\documentclass[12pt,preprint]{aastex}

\newcommand{\kev}{keV}

\newcommand{\xmm}{\textit{XMM-Newton}}

\newcommand{\fe}{Fe~K$\alpha$}
\newcommand{\etal}{et al.}

\shorttitle{A correlation between the accretion disk ionization state and Eddington
  ratio in AGNs}
\shortauthors{Ballantyne \etal}

\begin{document}

\title{A Correlation Between the Ionization State of the Inner
  Accretion Disk and the Eddington Ratio of Active Galactic Nuclei}


\author{D. R. Ballantyne\altaffilmark{1},
  J. R. McDuffie\altaffilmark{1} and J.\ S.\ Rusin\altaffilmark{1,2}}
\altaffiltext{1}{Center for Relativistic Astrophysics, School of Physics, Georgia
  Institute of Technology, 837 State Street, Atlanta, GA 30332-0430;
  david.ballantyne@physics.gatech.edu}
\altaffiltext{2}{South Cobb High School, 1920 Clay Road, Austell, GA 30106-2299}

\begin{abstract}
X-ray reflection features observed from the innermost regions of
accretion disks in Active Galactic Nuclei (AGNs) allow important tests
of accretion theory. In recent years it has been possible to use the
\fe\ line and reflection continuum to parametrize the ionization state of the irradiated inner accretion disk. Here, we collect 10
measurements of $\xi$, the disk ionization parameter, from 8 AGNs with
strong evidence for reflection from the inner accretion disk and
good black hole mass estimates. We find strong statistical evidence
($98.56$\% confidence) for a nearly linear correlation between $\xi$ and the AGN
Eddington ratio. Moreover, such a correlation is predicted by a simple
application of $\alpha$-disk accretion theory, albeit with a stronger
dependence on the Eddington ratio. The theory shows that there will be
intrinsic scatter to any correlation as a result of different black
hole spins and radii of reflection. There are several possibilities
to soften the predicted dependence on the Eddington ratio to allow a
closer agreement with
the observed correlation, but the current data does not allow for an
unique explanation. The correlation can be used to estimate
that MCG--6-30-15 should have a highly ionized inner accretion disk,
which would imply a black hole spin of $\sim 0.8$. Additional measurements of
$\xi$ from a larger sample of AGNs are needed to confirm the existence
of this correlation, and will allow investigation of the accretion
disk/corona interaction in the inner regions of accretion disks.
\end{abstract}


\section{Introduction}
\label{sect:intro}
It has been decades since it was realized that active galactic nuclei
(AGNs) must be powered by accretion onto supermassive black holes
\citep{lb69,rees84}. Yet, while our theoretical understanding of the
underlying physics of accretion disks has evolved tremendously over
that time \citep[e.g.,][]{nt73,ss73,pri81,bh98,balb03,hkb09}, there are
few direct measurements of the properties of AGN accretion disks
that can be used to test the theories. The difficulty in detecting AGN
accretion disks observationally arises naturally: they emit most of
their radiation in the ultraviolet and optical, which can be easily
absorbed and reprocessed by surrounding material and must also be
separated from the starlight from the surrounding galaxy. When
estimates of the disk continuum have been observationally determined
(typically from the outer part of the disk) they indicate broad
agreement with the concept that the accretion disk is an optically
thick, thermal emitter with a radially dependent temperature profile
\citep[e.g.,][]{kish05,kish08}. However, the few probes of the
accretion disk emission from close to the black hole show signs of
significant discrepancies with the theoretical
expectation\footnote{This should be contrasted with the situation for
accretions disks around stellar-mass black holes, whose continuum is
emitted at X-ray energies and is, in general, well matched by the
theoretical predictions \citep[e.g.,][]{dav06}. However, stellar mass
black holes also present several phenomenological problems (e.g., QPOs)
that are yet to be fully explained by accretion theory (see, e.g.,
\citealt{jp11} and references therein).}
\citep[e.g.,][]{kb99,dwb07,mor10}. Accretion disk theory also predicts
specific changes in the flow properties (e.g., density, temperature)
as a function of black hole mass and accretion rate \citep{ss73}, but
there are currently very few observational tests of these predictions
of the theory \citep[e.g.,][]{bon07,dwb07,mor10,dl11}.

A direct probe of the inner accretion disks of AGNs is through the
X-ray emission. All AGNs emit a fraction of their bolometric energy as
a hard X-ray power-law \citep[e.g.,][]{mu93,vf07}. The X-rays are
thought to originate in a hot and tenuous magnetically dominated corona
that sits above the surface of the disk
\citep{gal79,haa91,haa93,haa94}. Studies of this X-ray emission by
generations of X-ray observatories discovered that in many AGNs there
are spectral features superimposed on the power-law that could be
easily explained as reprocessing of the X-rays in nearby cold, dense
material \citep{pou90,np94}. The strongest
of these `X-ray reflection' features are the \fe\ emission line at
6.4~\kev, the associated Fe~K absorption edge at 7.1~\kev, and an overall
hardening of the spectrum at high energies due to electron scattering
of the X-ray photons \citep{lw88,gr88,gf91,mpp91}. As the inner
accretion disk likely subtends a large solid angle as seen from the
X-ray source, it may dominate the X-ray reflection spectrum, although,
depending on the source, contributions can arise from distant
material such as the broad line region, the dusty absorber or the
outer accretion disk \citep{nan06,shu10}. High signal-to-noise X-ray
spectral observations of some AGNs find a broadened, asymmetric \fe\
line. The shape of this line is strong evidence that the reflecting material is close
to the black hole and subject to strong relativistic effects
\citep{tan95,fab02,mill07}. Thus, the reflection spectra in these AGNs
can probe physical properties of the inner accretion disk
that would otherwise be extremely difficult to investigate by consideration of the
disk continuum. The highly penetrating X-ray emission can be studied
for nearly all AGNs and with very little contamination from the host
galaxy. The only difficulty is to identify and model the component of
the reflection spectrum that arises from the inner disk, as this will
often be a very weak and low contrast feature in the total spectrum
\citep{nan07,ball10,fero10}.

Once identified, the reflection spectrum from the inner accretion disk
can constrain the iron abundance of the accreting gas \citep{bfr02}, the
vertical density structure \citep{nkk00,brf01,btb04}, and its
ionization state, which is the focus of the current paper. The
spectral features imprinted on the reflection spectrum are a strong
function of the ionization state of the illuminated gas
\citep[e.g.,][]{rf93,rfy99,rf05,gk10}. In particular, the \fe\ line
transitions from a rest energy of 6.4~\kev\ to 6.7~\kev\ and then to
6.97~\kev\ as the iron ions become progressively more ionized. In
addition, the Fe~K absorption edge moves to higher energies and
Comptonization broadens out the natural width of the spectral
features. The ionization of lower Z elements also tends to flatten out
the reflection spectrum \citep{rf05,gk10}. All these effects are
potentially observable in high quality X-ray spectral data. Indeed,
evidence for ionization of the disk surface has been identified in the
X-ray spectra of several AGNs
\citep[e.g.,][]{bif01,fab04,long07,br09,nard11}.

The photoionization of the surface of the inner accretion disk by the
X-ray emitting corona can be parametrized by an ionization parameter,
$\xi = 4\pi F_{X}/n_{\mathrm{H}}$, where $F_{X}$ is the incident X-ray
flux and $n_{\mathrm{H}}$ is the hydrogen number density of the disk
surface. A measurement of $\xi$ can therefore provide information on
the density structure of the disk and/or the illuminating conditions
provided by the corona. Models of coronal generation predict that the
amount of accretion energy dissipated in the corona may be a function
of the accretion rate \citep{sr84,mf02,bp09}, and observations show that
the fraction of the bolometric luminosity released by an AGN in the
X-ray band decreases with increasing Eddington ratio
\citep[e.g.,][]{wang04,vf07}. These two results imply there should be
a dependence of $\xi$ on the accretion rate \citep[cf.,][]{br02},
which, if measured, can test how the density structure and corona
power varies with accretion rate. Earlier work by \citet{ith07}
indicated that $\xi$ increases with accretion rate, but those authors
were unable to fit the X-ray data with ionized reflection models to
measure $\xi$ at various Eddington ratios.

In this paper we measure the dependence of $\xi$ on AGN accretion rate
by compiling the best available results from the literature. The sources were selected to
have clear evidence of reflection from the inner accretion disk and
robust black hole mass estimates. The next section
describes the source selection in detail and presents the results of
our experiment which are then discussed in the context of accretion
and coronal models in Section~\ref{sect:discuss}. Our conclusions our
summarized in Section~\ref{sect:concl}.
 
\section{Source Selection and Results}
\label{sect:data}
The literature was searched to find the best available observationally
determined values of $\xi$. Sources selected for analysis must pass two criteria: first, the
central black hole mass must be estimated either from reverberation
mapping, or, if the radius-luminosity relationship \citep[e.g.,][]{kas05} is used, a
careful decomposition of the H$\beta$ line must be performed to obtain
an estimate of the virial motion of the broad-line region (BLR)
gas. Second, the X-ray spectrum of the object must have strong
evidence of reflection from the inner accretion disk (within
$10$~$r_g$, where $r_g=GM/c^2$ is the gravitational radius of a black
hole with mass $M$) in the form of a broad \fe\ line. AGNs where
relativistic effects are not needed to describe the line profile are
not included for analysis. These two requirements severely restricts
the number of possible AGNs that can be included in the sample, but
the objects that are included will provide the cleanest test of any
relationship between $\xi$ and the Eddington ratio with a minimal
amount of observational scatter. Finally, the AGNs must have a measurement of $\xi$
from fitting a relativistically blurred ionized reflection spectrum to
the \fe\ line and continuum. For consistency, this $\xi$ must be
estimated from the `reflionx' model \citep{rf05}. After searching
through the literature, it was found that 10 observations of 8
different AGNs satisfied the above criteria\footnote{Objects with
  relativistic \fe\ lines that
  failed to make the list because of an uncertain black hole mass
  include MCG--6-30-15, IRAS 13224-3809, 1H 0707-495 and Swift
  J2127.4+5654. Radio-loud AGNs with reverberation mapped black hole
  masses such as 3C 120, 3C 273 and 3C 390.3 were also
  not included to eliminate confusion due to emission from the
  relativistic jet.} (see
Table~\ref{table:litobjects}). The $2$--$10$~\kev\ bolometric
corrections, $\kappa_{\mathrm{X}}$, are found from the
measurements of \citet{vf09}, or, for three of the objects, are estimated by using the
observed relationship between $\kappa_{\mathrm{X}}$ and photon
index \citep{zz10}.

\begin{figure}
\plotone{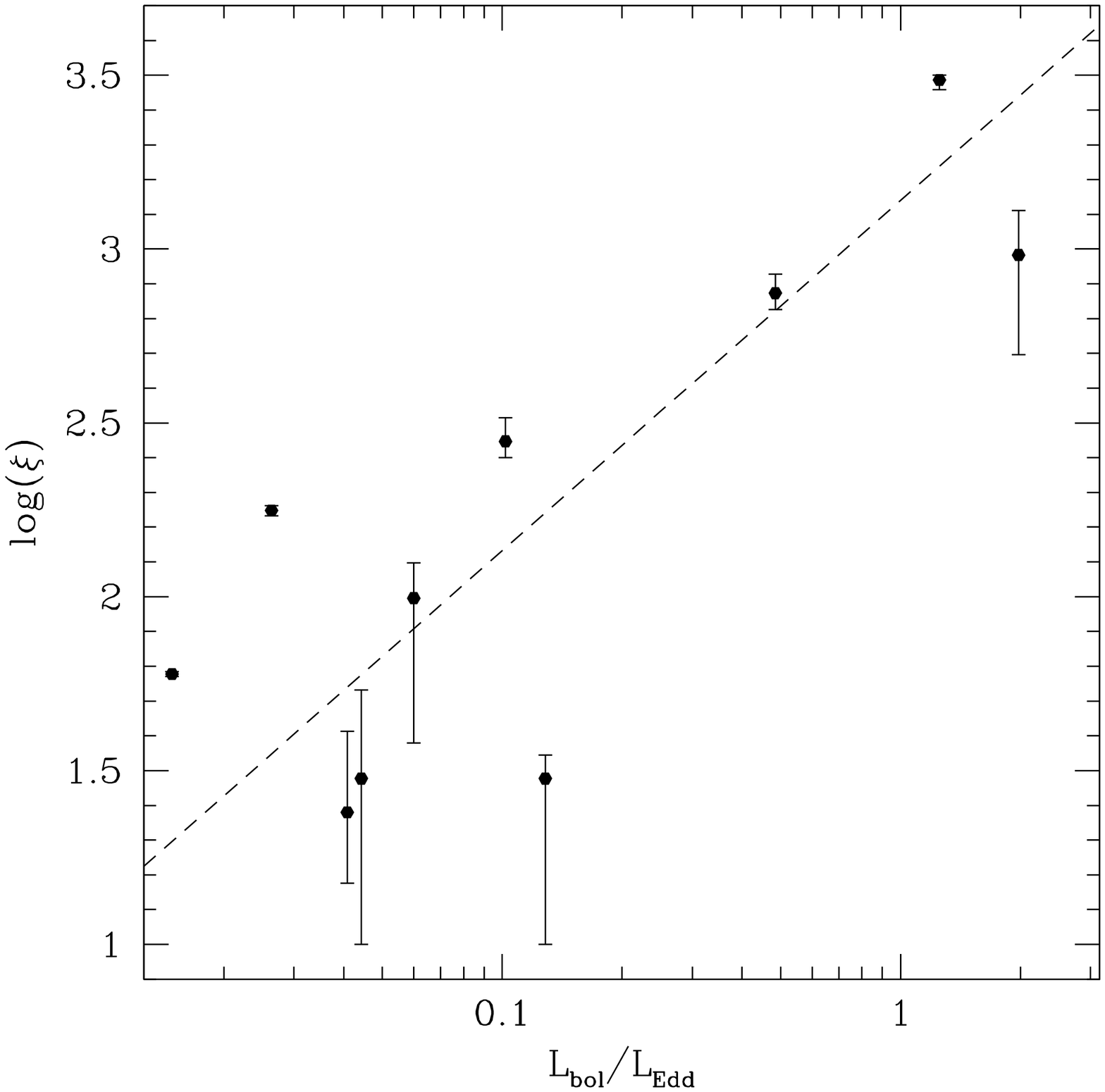}
\caption{A plot of $\log \xi$ versus
  $L_{\mathrm{bol}}/L_{\mathrm{Edd}}$ for the AGNs listed in
  Table~\ref{table:litobjects}. A significant correlation is observed with a
  Spearman rank correlation coefficient of 0.685 ($98.56$\% confidence
  level). The dashed line plots ordinary least-squares bisector fit to
  the data (Eq.~\ref{eq:fit}).}
\label{fig:data}
\end{figure}

Figure~\ref{fig:data} plots $\log \xi$ versus the Eddington ratio for
the objects listed in Table~\ref{table:litobjects}, and shows a clear
correlation between the two parameters (the simple linear correlation
coefficient is $r=0.80$). The Spearman rank correlation coefficient is
$0.685$ and gives a $t$-value of $2.658$, which corresponds to a
$98.56$\% confidence level for 8 degrees of freedom. Thus, the
observed correlation has strong statistical significance. However, the
detection of the correlation depends on only three objects with
$(L_{\mathrm{bol}}/L_{\mathrm{Edd}})> 0.2$, and, as seen in
Table~\ref{table:litobjects}, those AGNs have only estimates of the
bolometric correction. Removing those three objects from the sample
eliminates the correlation ($r=0.04$); it is therefore crucial to
obtain better
measurements of bolometric corrections for high accretion rate AGNs in
order to confirm the existence of the relationship between $\xi$ and
$L_{\mathrm{bol}}/L_{\mathrm{Edd}}$.

To estimate the functional relationship between these quantities, we
follow the advice of \citet{isobe90} who recommend the ordinary
least-squares bisector method. This calculation finds the following,
nearly linear, relationship between $\xi$ and the Eddington ratio:
\begin{equation}
\label{eq:fit}
\log \xi = (1.008 \pm 0.162) \log (L_{\mathrm{bol}}/L_{\mathrm{Edd}}) +
  (3.14 \pm 0.164).
\end{equation}
This fit is shown as the dashed line in Fig.~\ref{fig:data}, and is
formally consistent with the slope ($0.805$) and intercept ($2.95$)
calculated from a simple least-squares fit. If the observed errors in $\xi$ are
included than this line gives a reduced $\chi^2 >> 1$; this can be
reduced to $\sim 1$ if all the errors in $\log \xi$ are increased to
$\pm 0.4$. However, as discussed below, substantial scatter in this
relationship is likely to arise naturally, so a reduced $\chi^2 \sim
1$ would not be expected. 
 
\section{Discussion}
\label{sect:discuss}

\subsection{Physical Interpretation}
\label{sub:interp}
Figure~\ref{fig:data} clearly shows that AGNs that are
accreting at larger fractions of their Eddington rate exhibit more
ionized inner accretion disks. As mention in Sect.~1, such a
relationship might be expected depending on how accretion energy is
dissipated in the corona at different accretion rates. In this
Section, we use simple $\alpha$-disk accretion theory to explore the
possible physical origins of this observed relationship.

As the $\xi$ measurements were all chosen to arise from the innermost
regions of the accretion disk, radiation pressure will likely dominate
the support of the accretion flow. Likewise, relativistic effects will
also be non-negligible. Finally, it will also be necessary to include
the effects of a non-zero (and potentially variable) fraction of
accretion energy dissipated in the corona, $f$. The ionization
parameter can be approximated as $\xi
\approx m_p L_{X}/H^2 \rho$, where $m_p$ is the mass of a proton,
$L_{X}$ is the total X-ray luminosity incident on the accretion disk
which has a gas density $\rho$, and $H$ is the distance from the X-ray
source in the corona to the reflecting region of the disk. As there is
no complete understanding of the structure of the corona close to the
black hole, for simplicity we will assume a geometrically thick corona
so that $H/R = 1$, where $R$ is the radial distance along the disk. Theoretical arguments by \citet{bp09} show that the survival of a
magnetically dominated corona will require large scale magnetic fields, so a thick
corona above the inner disk may not be unrealistic (see also
\citealt{mf01}). With this assumption, Appendix~\ref{sect:app1} shows that for a
radiation pressure supported disk, 
\begin{equation}
\xi \approx (4.33\times 10^9) \left ( {\eta \over 0.1} \right )^{-2}
\left ({\alpha \over 0.1} \right ) \left ( {L_{\mathrm{bol}} \over
  L_{\mathrm{Edd}} } \right )^3 \left ( {R \over r_g} \right )^{-7/2}
R_z^{-2} R_T^{-1} R_R^3 f(1-f)^3  \ \mathrm{erg\ cm\ s^{-1}},
\label{eq:radxi}
\end{equation}
where $\eta$ is the radiative efficiency of the disk, $\alpha$ is the
viscosity parameter \citep{ss73}, and $(R_R, R_z, R_T)$ encompasses the
general relativistic effects and are
dimensionless functions of $a_{\ast}$, the dimensionless black hole spin, and $(R/r_g)$
\citep[e.g.,][]{nt73}. This estimation of $\xi$ is nicely independent of the
central black hole mass. Initially, we assume constant values of
$\alpha=0.1$ and $\eta=0.089$ \citep{dl11}, so a model curve of $\xi$
versus $(L_{\mathrm{bol}}/L_{\mathrm{Edd}})$ can be described
by only three parameters: $a_{\ast}$, $R/r_g$ and $f$.

\begin{figure}
\plotone{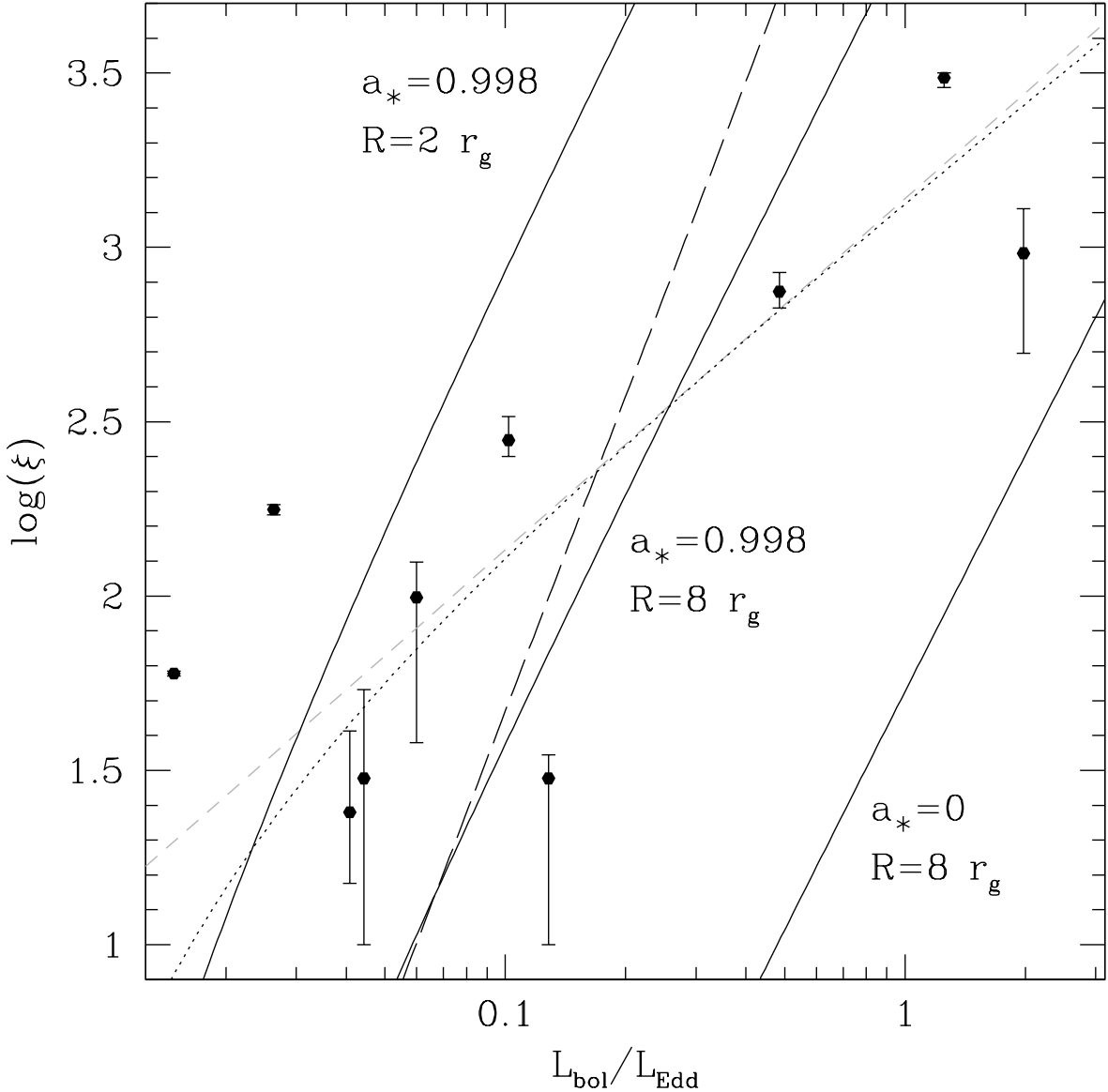}
\caption{As in Fig.~\ref{fig:data}, but overlaid with model curves
  derived from Eq.~\ref{eq:radxi}. To guide the eye, the gray dashed line show the least
  squares bisector fit from Eq.~\ref{eq:fit} and
  Fig.~\ref{fig:data}. The solid lines plot the predicted
  $\xi$--$(L_{\mathrm{bol}}/L_{\mathrm{Edd}})$ relationship from
  Eq.~\ref{eq:radxi} for different combinations of $a_{\ast}$, the
  black hole spin, and $R$, the radius of peak reflection. These
  particular models assume $f \propto
  (L_{\mathrm{bol}}/L_{\mathrm{Edd}})^{-0.77}$ \citep{sr84}. The long
  dashed line plots the $a_{\ast}=0.998$, $R=8$~$r_g$ model if a
  constant $f=0.45$ is assumed \citep{vf07}. The dotted black line shows the
  $a_{\ast}=0$, $R=8$~$r_g$ model when the disk density has been
  reduced by a factor that is proportional to
  $(L_{\mathrm{bol}}/L_{\mathrm{Edd}})^{-1.3}$. This factor ranges
  from $\sim 10^4$ for $L_{\mathrm{bol}}/L_{\mathrm{Edd}}=0.01$ to
  $25$ at $L_{\mathrm{bol}}/L_{\mathrm{Edd}}=1$. As described in the
  text, this is only one out of several possibilities to bring the
  slope of the model into agreement with the observed relationship.}
\label{fig:fits}
\end{figure}

Figure~\ref{fig:fits} plots several models calculated from
equation~\ref{eq:radxi} and compares them against the data and the
least-squares bisector fit from Fig.~\ref{fig:data}. There are three
sets of curves for three different $(a_{\ast}$, $R/r_g)$ pairs. As
described in detail below, the solid lines and the long dashed line
differ in the treatment of $f$. Fig.~\ref{fig:fits} is remarkable in that the simple
$\alpha$-disk model of Eq.~\ref{eq:radxi} predicts reasonable values of
$\xi$ for realistic values of the parameters; there is no reason for
this to be expected \emph{a priori}. This basic level of agreement
provides confidence that Eq.~\ref{eq:radxi} can be used to explore the
physical reasons behind the observed correlation between $\xi$ and
$(L_{\mathrm{bol}}/L_{\mathrm{Edd}})$.

There can be a wide range of expected values of $\xi$ for a given
value of $L_{\mathrm{bol}}/L_{\mathrm{Edd}}$ depending on the value of
the black hole spin and the radius which dominates the reflection
signal. As these values will likely vary from object to
object, and the latter may vary from observation to observation, a
fair amount of natural scatter will occur in the observed
relationship. In fact, this scatter may be a significant contribution
to the observed slope of the relationship; however, the scatter is not
so significant as to destroy the correlation. This fact argues that
AGNs within this range of Eddington ratios are not uniformly
distributed over $a_{\ast}$ and $R$, but are confined to a relatively
small range in these parameters. Clearly, more ionized reflection fits
in a wider range of AGNs is needed to better understand the intrinsic
scatter in this relationship.

\citet{vf07} make use of their measured AGN X-ray bolometric
corrections to estimate that the coronal fraction $f \approx 0.45$ for
low Eddington rate AGNs. The long dashed line in Fig.~\ref{fig:fits}
shows the predicted $\xi$--$(L_{\mathrm{bol}}/L_{\mathrm{Edd}})$
relationship from eq.~\ref{eq:radxi} for this value of $f$ when $a_{\ast}=0.998$ and
$R=8$~$r_g$. This line has a slope of $3$, as predicted from
eq.~\ref{eq:radxi}, much steeper than the measured slope of $\sim 1$
(eq.~\ref{eq:fit}). However, \citet{vf07} also found that $f$ is lower
for high Eddington ratio sources, dropping to $\sim 0.11$. The
anti-correlation of $f$ with Eddington ratio is a natural consequence
of the well established diminishing of relative X-ray power in high
luminosity quasars \citep[e.g.,][]{steff06}. Models of accretion disk corona predict
different dependencies of $f$ on the Eddington ratio, with \citet{wang04}
providing a comparison with observational estimates. The steepest
dependence of $f$ with $(L_{\mathrm{bol}}/L_{\mathrm{Edd}})$ is
provided by the model of \citet{sr84} which finds $f \propto
  (L_{\mathrm{bol}}/L_{\mathrm{Edd}})^{-0.77}$. Inserting this
dependence into eq.~\ref{eq:radxi}, and normalizing so that $f=0.45$ at
$(L_{\mathrm{bol}}/L_{\mathrm{Edd}})=0.01$, yields the solid lines in
figure~\ref{fig:fits} which have slopes of $\approx 2.3$, still
significantly steeper than the observed slope.

Although the intrinsic scatter is likely to contribute to the mismatch
between the observed and predicted slope, it is interesting to consider
what other physical mechanisms may be operating that can bring the
model closer to the observed correlation. There are numerous
possibilities, but, as outlined below, not all are likely physical
reasonable.

\subsubsection{Coronal Fraction} 
If $f \propto
  (L_{\mathrm{bol}}/L_{\mathrm{Edd}})^{-2}$ then the slope of the
  model is approximately equal to the observed one. However, this
  dependence is so strong that, if normalized to $0.45$ at
  $(L_{\mathrm{bol}}/L_{\mathrm{Edd}})=0.01$, $f$ drops to 1\% at an
  Eddington ratio of only 0.07. Such a strong dependence of $f$ would
  be very difficult to reconcile with the observed X-ray luminosities of luminous quasars. 

\subsubsection{Gas Density} 
Perhaps the simplest change to the model is to alter the gas
  density of the disk. The value of $\rho$ used in calculating $\xi$
  is an approximate, vertically-averaged value of the gas density, and
  may not be entirely accurate for the outer layers of the disk that
  is being illuminated by the X-rays. Furthermore, the X-ray heating
  may cause the disk scale height to increase and thus decreasing the
  density \citep[e.g.,][]{nkk00,brf01}. As there is more accretion
  power emitted as X-rays at low Eddington ratios, then its plausible
  that this density correction could be dependent on
  $(L_{\mathrm{bol}}/L_{\mathrm{Edd}})$. The dotted line in
  figure~\ref{fig:fits} shows the $a_{\ast}=0$, $R=8$~$r_g$ model
  after reducing the density with a factor that varies as
  $(L_{\mathrm{bol}}/L_{\mathrm{Edd}})^{-1.3}$. The reduction in
  density ranges by a factor of $\sim 10^4$ at
  $(L_{\mathrm{bol}}/L_{\mathrm{Edd}})=0.01$ to $25$ for an Eddington
  ratio of unity. 

There are problems with this solution, however. First, it can most
easily be applied to low spin black holes. Higher spin black holes,
such as the $a_{\ast}=0.998$ models plotted in figure~\ref{fig:fits}
would require a \emph{denser} disk at high Eddington ratios to bring
the model into agreement with the data. In addition, reflection calculations of disks which
do adjust into hydrostatic balance and have low density surfaces
predict that ionized \fe\ lines are common \citep{br02} unless the
disk is very weakly illuminated. It is not clear that this is
consistent with the observed variation in coronal fractions, where
more accretion power is dissipated in the corona at low Eddington
ratios and the ionization parameter is observed to be small. Both
these problems may be mitigated if processes within the disk (e.g.,
photon bubble or Parker instabilities) transported denser material to the surface
\citep{btb04,bhk07}. More simulations investigating the density structure of accretion
disk photospheres are needed to quantify the possible corrections to
the $\alpha$ disk densities.

\subsubsection{Radiative Efficiency}
According to equation~\ref{eq:radxi}, $\xi \propto \eta^{-2}$, so if
the radiative efficiency of accretion disks was proportional
$(L_{\mathrm{bol}}/L_{\mathrm{Edd}})^{7/10}$ then, after including the
effects of the variable $f$, the slope of the
model would be more in line with the observed correlation. Of course,
$\eta$ is a measure of the total binding energy available to be radiated
away in the accretion flow and thus depends on the black hole spin and
the innermost radius of the disk \citep[e.g.,][]{kro99}. The maximal values of
$\eta$ are limited to be 0.31 (0.038) for maximally spinning black
holes that are co-rotating (counter-rotating) with respect to their accretion
disks [see \citet{dl11} and references therein]. Additional physics can
alter these expectations, however; for example, magnetohydrodynmical
effects may provide an extra torque to the gas within the traditional
innermost stable orbit and increase the efficiency above the
theoretical upper-limit \citep[e.g.,][]{ak00}. Alternatively, accretion energy may be
advected inwards or used to drive outflows which can reduce the
efficiency. Observationally, arguments based on the work of \citet{sol82} have
shown that the average efficiency of accreting black holes over their
accretion history is $\ga 0.1$ \citep[e.g.,][]{yt02,elv02,marc04,barg05}. Recently, \citet{dl11}
estimated $\eta$ for 80 PG quasars and found an average value of $\sim
0.1$ (with a wide scatter) and no strong correlation with the
Eddington ratio. Indeed, implementing the
$\eta \propto (L_{\mathrm{bol}}/L_{\mathrm{Edd}})^{7/10}$ dependence
into equation~\ref{eq:radxi} finds that it can only hold over a factor
of $\sim 20$ in Eddington ratio before reaching the theoretical
limits.

Interestingly, \citet{dl11} find a positive correlation between $\eta$ and
black hole mass, $\eta \propto M^{1/2}$. Thus, an alternative way to
reduce the slope of the $\xi$--$(L_{\mathrm{bol}}/L_{\mathrm{Edd}})$
relationship is with $(M/M_{\odot}) \propto
(L_{\mathrm{bol}}/L_{\mathrm{Edd}})^{4/3}$. Again, such a strong
relationship is not observed \citep{koll06,se10,dl11}, nor is it even hinted at with the
data in Table~\ref{table:litobjects} where only a slight negative
correlation is possible ($r=-0.44$).

\subsubsection{Disk Viscosity}
The models shown in fig.~\ref{fig:fits} assume a constant
$\alpha=0.1$, but the slope of the predicted relationship could be
reduced to $\sim 1$ if  $\alpha \propto
(L_{\mathrm{bol}}/L_{\mathrm{Edd}})^{-1.3}$. Observational estimates of
$\alpha$ typically find $\alpha \ga 0.1$ \citep[e.g.,][]{kpl07}, but numerical
simulations of accretion flows typically find values an order of
magnitude lower \citep[e.g.,][]{dsp10}. The strong relationship between $\alpha$ and
the Eddington ratio required to account for the observed
$\xi$--$(L_{\mathrm{bol}}/L_{\mathrm{Edd}})$ relationship results in
either a large fraction of low Eddington AGNs with $\alpha \sim 1$, or
that most rapidly accreting AGN have $\alpha < 0.01$. As the $\alpha$
parameter is at best a parametrization of our ignorance of the
accretion process, it is difficult to assess the validity of such a
relationship between $\alpha$ and accretion rate without guidance from
simulations that explore disk viscosity as a function of $\dot{M}$.

\subsubsection{Black Hole Spin}
Figure~\ref{fig:fits} shows that the black hole spin has an important
impact on the predicted values of $\xi$, and that the slope of the
predicted curves could be reduced if $a_{\ast}$ was a decreasing
function of $(L_{\mathrm{bol}}/L_{\mathrm{Edd}})$, an idea
that has some theoretical justification. In particular, some models of
black hole growth predict that only the largest black holes that have
stopped rapidly accreting have developed high spins, while the lowest
mass black holes which may still undergo rapid accretion will
generally have low values of $a_{\ast}$ \citep[e.g.,][]{fan11}. That is, the high
Eddington ratio AGNs would be lower mass black holes with small spins,
and the low Eddington ratio AGNs would be high mass black holes with
large spins. This scenario is also a possible explanation for the
correlation between $\eta$ and black hole mass found by
\citet{dl11}. Although the objects in our sample do not
present strong evidence for a correlation between black hole mass and
Eddington ratio, we assumed that $a_{\ast}
\propto (L_{\mathrm{bol}}/L_{\mathrm{Edd}})^x$ and searched for an $x$
that could reduce the predicted slope to match the observations. The
minimum slope found was $1.4$ for $x=-0.11$ and a reflecting radius
very close to the innermost stable circular orbit. Thus, a dependence
of spin with Eddington ratio cannot, \emph{on its own}, reproduce the
observed correlation between $\xi$ and
$(L_{\mathrm{bol}}/L_{\mathrm{Edd}})$. 

\subsubsection{Other Options}
There are other physical effects neglected in equation~\ref{eq:radxi}
that could influence the observed $\xi$. For example, if the X-ray
source is close to the black hole, gravitational light bending
\citep[e.g.,][]{mf04} will
increase the irradiation of the disk and thus increase $\xi$. If this
was more common, or more effective, at smaller
$(L_{\mathrm{bol}}/L_{\mathrm{Edd}})$ (as might be expected given the
larger values of $f$), then this effect could reduce the slope of
the model to be closer to the observed value. However, if the disk is
more highly illuminated than this would also increase the strength of
the reflection spectrum and increase the equivalent width (EW) of the broad
\fe\ lines. Iron lines with large (i.e., several hundred eV) EWs are rare in
Seyfert galaxies \citep{nan07,fero10} which argues that strong light
bending is not a common phenomenon (see also
\citealt{ball10}). Alternatively, it is possible some fraction of the
accretion energy dissipated in the corona is used to drive an outflow
\citep[e.g.,][]{mf02}. This would have two important effects for the
predicted value of $\xi$: first, this would reduce the energy
available to illuminate the accretion disk, and, second, the X-rays
emitting region may be moving rapidly away from or toward the disk and thus
significantly decrease or increase the irradiating flux \citep{bel99b}. As
$f$ is likely to depend on Eddington ratio, then it is certainly
plausible that such effects may also be stronger over a certain range
of $(L_{\mathrm{bol}}/L_{\mathrm{Edd}})$. More generally, it is possible (perhaps likely) that all of the effects considered in
this section are in play at some level in real AGNs. Measurements
of $\xi$ from the inner regions of accretion disk are needed from
several more AGNs with reverberation mapped black hole masses to
further explore the nature of this correlation. 

\subsection{Application to MCG--6-30-15}
\label{sub:app}
One potentially important application of this correlation is that it
may help distinguish between two competing interpretations of an
observed X-ray spectrum. For example, MCG--6-30-15 exhibits one of the
broadest and strongest \fe\ lines ever observed and has been analyzed
many times since the line was first detected by \textit{ASCA}
\citep[e.g.,][]{tan95,bf01,wil01,fab02,bvf03,rey04,br06,min07}. When modeled as one powerful relativistic line, reflection fits to the
spectra of MCG--6-30-15 yield $\log \xi \la 2$ and require a large
iron abundance \citep[e.g.,][]{br06,min07}. However, equally good fits to the spectra can
be found if the red wing of the line arises from highly ionized Fe
close to the black hole and the neutral 6.4~\kev\ core arises from
much larger distances on the disk \citep{bf01,bvf03,rey04}. None of the AGNs
in Table~\ref{table:litobjects} show evidence for a strong second
reflector from the accretion disk that could bias the measurement of
the ionization parameter and the resulting correlation. Thus, this
degeneracy is unlikely to be common, but, given the importance of
MCG--6-30-15 to the field, it is interesting to use the correlation
observed in Fig.~\ref{fig:data} to provide additional information on
the likely ionization state of the MCG--6-30-15 accretion disk.

Unfortunately, the black hole mass of this source is not well
constrained with a current best estimate of $4\times 10^6$~M$_{\odot}$
\citep{mch05}. The flux of MCG--6-30-15 when it was observed in a deep minimum
state ($2.3\times 10^{-11}$~erg~cm$^{-2}$~s$^{-1}$; \citealt{wil01}) is used to
estimate a lower limit to $\log \xi$. As there is no direct
measurement of $\kappa_X$, we utilize the photon index of MCG--6-30-15
in the deep minimum state ($\Gamma=1.8$; \citealt{rey04}) and again make use of the correlation
between $\kappa_X$ and $\Gamma$ described by \citet{zz10} which
results in $\kappa_X \approx 24$. Combining these measurements results
in $L_{\mathrm{bol}}/L_{\mathrm{Edd}}=0.18$ which when inserted into
Eq.~\ref{eq:fit}, and taking the extremal values of the slope and
intercept, yields $\log \xi > 2.1$. This lower limit suggests that the
accretion disk of MCG--6-30-15 is most likely to be significantly
ionized.

Interestingly, when the broad \fe\ line of MCG--6-30-15 is fit by
a highly ionized reflector, the ionization parameter is usually $\log
\xi > 3.5$ \citep{bvf03,rey04}, and such a highly
ionized line is naturally broader due to Comptonization
\citep[e.g.,][]{rfy99}. Thus, an ionized accretion disk in
MCG--6-30-15 would reduce the value of $a_{\ast}$
required to explain the width of the \fe\ line \citep[e.g.,][]{rey04,br06}. However, the fits of \citet{bvf03} found that the ionized reflector
was constrained to a radius of $R \sim
5$~$r_g$, and therefore a non-zero black hole spin will be necessary. We can
be more quantitative by combining the measured $\xi$ and $R$ from
spectral fits with the predicted
$\xi$--$(L_{\mathrm{bol}}/L_{\mathrm{Edd}})$ correlation. An
ionization parameter of $\log
\xi > 3.5$ would be larger than expected for the correlation seen in Fig.~\ref{fig:data}, but scatter due to the various values of
$a_{\ast}$ and $R$ will bring objects above that line. According to
Equation~\ref{eq:radxi}, if $\log \xi > 3.5$ at $R \sim 5$~$r_g$ for
Eddington ratios $\sim 1$ \citep{min07}, then $a_{\ast} \sim
0.8$. Thus, a rapidly spinning black hole is still necessary for
MCG--6-30-15, but a very large value is not required if the inner disk
is ionized.

\section{Conclusions}
\label{sect:concl}
In this paper we have presented evidence from the literature that the ionization state of
the inner accretion disk in AGNs is correlated with the Eddington
ratio of the accretion flow. Very conservative criteria were employed
to select the data from the literature; namely,
there must be strong evidence for reflection from within $10$~$r_g$,
the reflection spectrum must have been fit with the latest ionized
disk models, and there must be a high quality estimate of the central
black hole mass. Despite these steps, the correlation relies on
estimates of the X-ray bolometric corrections for the high Eddington
rate sources. Therefore, precise and accurate measurements of
bolometric corrections (especially of rapidly accreting AGNs) are
needed to both confirm the existence of this correlation and to better
define its intrinsic scatter.

The existence of a relationship between $\xi$ and
$L_{\mathrm{bol}}/L_{\mathrm{Edd}}$ is predicted by simple $\alpha$
disk accretion theory, although with a slope that is steeper than what
is observed. It is not possible to determine an unique explanation that
can bring the two slopes into agreement, although a Eddington
ratio dependent decrease in the disk density (possibly related to the
changing coronal power) presents the simplest possibility. If the
correlation can be better defined by future observations than it may
lead to insight into the physics of the energy flow between the disk
and X-ray emitting corona.

Finally, we showed that the
$\xi$--$(L_{\mathrm{bol}}/L_{\mathrm{Edd}})$ correlation will also be
useful in distinguishing between different interpretations of AGN X-ray
spectra. As an example, given the current estimate of the black hole
mass of MCG--6-30-15, the correlation predicts that the
accretion disk of MCG--6-30-15 should be significantly ionized with
$\log \xi > 2.1$. This result provides evidence that the strong line
observed from this object may arise from two distinct reflecting
regions with different ionization states, rather than one low-$\xi$ area
with a very high Fe abundance. The high values of $\xi$
measured in this interpretation require $a_{\ast} \sim 0.8$,
reducing the value of the spin necessary to explain the width of the
red wing.

\acknowledgments
We thank the referee, Dr.\ Chris Reynolds, for a helpful report that
improved the paper. This work was supported in part by NSF award AST
1008067 to DRB.

\appendix
\section{The Ionization Parameter in a Radiation Pressure Dominated
  Disk}
\label{sect:app1}
To estimate the ionization parameter of an accretion disk illuminated
at a radius $R$ by a X-ray luminosity $L_X$, we re-write $\xi=4\pi
F_{X}/n_{\mathrm{H}}$ as $\xi \approx m_p L_{X}/H^2 \rho$, assume $H/R
= 1$ (see Sect.~\ref{sub:interp}), and use
the density of a radiation pressure dominated disk provided by
\citep{kro99}:
\begin{equation}
\rho = (2.23\times 10^{-6}) \left ( {\eta \over 0.1} \right )^2 \left
( {\alpha \over 0.1} \right )^{-1} \left ( {M \over M_{\odot}} \right
)^{-1} \left ( {L_{\mathrm{bol}} \over L_{\mathrm{Edd}}} \right )^{-2}
\left ( {R \over r_g} \right )^{3/2} R_z^2 R_T R_R^{-3} (1-f)^{-3}\ \mathrm{g\
  cm^{-3}},
\label{eq:app1.1}
\end{equation}
where the correction due to a non-zero coronal dissipation fraction
$f$ has been included from the results of \citet{sz94}. The
relativistic corrections $(R_R, R_z, R_T)$ are simple, but lengthy, analytic functions
of $a_{\ast}$, the dimensionless black hole spin, and the disk radius
$(R/r_g)$ and can be found elsewhere \citep[e.g.,][]{kro99}. 

Finally, it is assumed that that the X-ray luminosity is powered
entirely by the fraction of the accretion energy that is dissipated
in the corona; i.e., $L_{X} = fL_{\mathrm{bol}}$. That is, none of the
coronal energy is used to launch an outflow or jet, which is a
reasonable assumption given the relatively high Eddington ratios of the objects making up the sample
\citep[e.g.,][]{mf02}. Combining this assumption with
eq.~\ref{eq:app1.1} yields equation~\ref{eq:radxi}.

\section{The Ionization Parameter in a Gas Pressure Dominated
  Disk with Electron Scattering Opacity}
\label{sect:app2}
Although not likely applicable to the inner disk reflection observed
in the AGNs considered here, for
completeness we also derive an estimate for $\xi$ for the case of an
illuminated gas pressure dominated disk (with electron scattering
opacity). In this case, the gas density is \citep{kro99}:
\begin{equation}
\rho = (239) \left ( {\eta \over 0.1} \right )^{-2/5} \left
( {\alpha \over 0.1} \right )^{-7/10} \left ( {M \over M_{\odot}} \right
)^{-7/10} \left ( {L_{\mathrm{bol}} \over L_{\mathrm{Edd}}} \right )^{2/5}
\left ( {R \over r_g} \right )^{-33/20} R_z^{1/2} R_T^{7/10} R_R^{-3/10} (1-f)^{-3/10}\ \mathrm{g\
  cm^{-3}},
\label{eq:app2.1}
\end{equation}
where again the correction due to dissipation in the corona was
taken from the work of \citet{sz94}. Then, with the same approximation
as before for the production of the ionizing X-rays (i.e.,
$L_X=fL_{\mathrm{bol}}$), the ionization parameter is:
\begin{equation}
\xi \approx (4.03) \left ( {\eta \over 0.1} \right )^{2/5}
\left ({\alpha \over 0.1} \right )^{7/10} \left ( {M \over M_{\odot}} \right
)^{-3/10} \left ( {L_{\mathrm{bol}} \over L_{\mathrm{Edd}} } \right
)^{3/5} \left ( {R \over r_g} \right )^{-7/20} R_z^{-1/2} R_T^{-7/10}
R_R^{3/10} f(1-f)^{3/10}  \ \mathrm{erg\ cm\ s^{-1}}.
\label{eq:app2.2}
\end{equation}
The high density of a gas pressure dominated disk predicts a very low
$\xi$ for supermassive black holes, and with a much slower dependence
on the Eddington ratio than the radiation pressure dominated
disk.

\clearpage

\clearpage

\clearpage

\begin{deluxetable}{llcccccccccc}
\tabletypesize{\footnotesize}
\rotate
\tablewidth{0pt}
\tablecaption{\label{table:litobjects} Details of AGN Sample}
\tablecolumns{12}
\tablehead{
\colhead{Source} & \colhead{Type} & \colhead{$z$} & \colhead{$\xi$} & \colhead{$\xi$ Ref.} & \colhead{$L_{\mathrm{X}}$} & \colhead{$L_{\mathrm{X}}$ Ref.} & \colhead{$\kappa_{\mathrm{X}}$} &
  \colhead{$\kappa_{\mathrm{X}}$ Ref.} & \colhead{$M_{\mathrm{BH}}$} &
  \colhead{$M_{\mathrm{BH}}$ Ref.} &
  \colhead{$L_{\mathrm{bol}}/L_{\mathrm{Edd}}$} \\
\colhead{\phantom{i}} & \colhead{\phantom{i}} & \colhead{\phantom{i}}
& \colhead{(erg cm~s$^{-1}$)} & \colhead{\phantom{i}} & \colhead{(erg
  s$^{-1}$)} & \colhead{\phantom{i}} & \colhead{\phantom{i}} &
\colhead{\phantom{i}} & \colhead{(M$_{\odot}$)} &
\colhead{\phantom{i}} & \colhead{\phantom{i}}}
\startdata
Mrk 766 & NLS1 & 0.0129 & 961$^{+329}_{-465}$ & BR09 & $7.6\times 10^{42}$ & N07 & 57
  & ZZ10 & $1.7\times 10^6$ & W10 & 2 \\
NGC 3783 & Sy 1.5 & 0.0097 & 30$^{+24}_{-20}$ & BR09 & $1.1\times 10^{43}$ & N07 & 15 &
  VF09 & $2.8\times 10^7$ & W10 & 0.04 \\
NGC 4051 & NLS1 & 0.0023 & 99$^{+26}_{-61}$ & BR09 & $2.0\times 10^{41}$ & N07 & 67 &
  VF09 & $1.7\times 10^{6}$ & D10 & 0.06 \\
Ark 120 & Sy 1 & 0.0327 & 30$^{+5}_{-20}$ & BR09 & $9.4\times 10^{43}$ & N07 & 25 &
  VF09 & $1.4\times 10^{8}$ & W10 & 0.13 \\
Ark 120 & --- & --- & 280$^{+47}_{-29}$ & N11 & $7.5\times 10^{43}$ & P11 & --- &
  --- & --- & --- & 0.10 \\
Fairall 9 & Sy 1.2 & 0.047 & 24$^{+17}_{-9}$ & P11 & $1.3\times 10^{44}$ & P11 & 10.5 &
  VF09 & $2.6\times 10^{8}$ & P04 & 0.04 \\
Mrk 79\tablenotemark{a} & Sy 1.2 & 0.022 & 177$^{+6}_{-6}$ & G11 &
$1.7\times 10^{43}$ & G11 & 10.5 & VF09 & $5.3\times 10^{7}$ & P04 & 0.03 \\
Mrk 79\tablenotemark{b} & --- & --- & 60$^{+1}_{-1}$ & G11 & $9.5\times 10^{42}$ & G11
& --- & --- & --- & --- & 0.01 \\
I Zw 1 & NLS1 & 0.0611 & 3060$^{+107}_{-186}$ & G07 &
$4.3\times 10^{43}$ & G07 & 76 & ZZ10 & $2.0\times 10^{7}$ & M10 & 1.3 \\
Mrk 478\tablenotemark{c,}\tablenotemark{d} & NLS1 & 0.079 &
746$^{+100}_{-76}$ & Z08 & $2.3\times 10^{43}$ & G06 & 86 & ZZ10 &
$3.2\times 10^{7}$ & M10 & 0.48 \\
\enddata
\tablecomments{Details compiled from the literature of AGNs with
  high quality black hole masse estimates that also have X-ray spectra
  that exhibit clear evidence for a relativistic \fe\ line and have
  been fit with the `reflionx' ionized disk model of
  \citet{rf05}. $L_{\mathrm{X}}$ denotes the intrinsic rest-frame
  $2$--$10$~keV luminosity of the AGN, $M_{\mathrm{BH}}$ is the mass
  of the black hole, and $L_{\mathrm{bol}}$ is the estimated
  bolometric luminosity. The Eddington luminosity,
  $L_{\mathrm{Edd}}$, is defined as $L_{\mathrm{Edd}}= 1.3\times
  10^{38} (M_{\mathrm{BH}}/M_{\odot})$~erg~s$^{-1}$.
\textit{References}: BR09: \citet{br09}, N07: \citet{nan07}, ZZ10:
\citet{zz10}, W10: \citet{woo10}, VF09: \citet{vf09}, D10:
\citet{denney10}, P11: \citet{patrick11}, N11: \citet{nard11}, P04:
\citet{peterson04}, G11: \citet{gallo11}, G07: \citet{gallo07}, M10:
\citet{marziani10}, Z08: \citet{z08}, G06: \citet{gallo06}.}
\tablenotetext{a}{data taken from the long \textit{Suzaku} observation}
\tablenotetext{b}{data taken from the XMM6 observation}
\tablenotetext{c}{data taken from longest \xmm\ observation}
\tablenotetext{d}{\citet{gallo06} listed the $2.5$--$10$~\kev\ flux, so
$L_{\mathrm{X}}$ will be slightly underestimated}
\end{deluxetable}

\end{document}